\documentclass[%
jmp,
 amsmath,amssymb,
preprint,%
]{revtex4-2}

\usepackage{mathtools,autobreak,stmaryrd}
\usepackage[usenames,dvipsnames,svgnames,table]{xcolor}

\usepackage{graphicx,multirow}
\usepackage[caption=false,font=Large]{subfig}
\usepackage{dcolumn}
\usepackage{bm}

\usepackage[utf8]{inputenc}
\usepackage[T1]{fontenc}
\usepackage{mathptmx}
\usepackage{etoolbox}

\usepackage{amsthm}
\newtheorem*{theorem*}{Theorem}
\newtheorem*{corollary*}{Corollary}
\newtheorem*{proof*}{Proof}

\newcommand{\hatx}{\hat{x}}\newcommand{\hatp}{\hat{p}}

\DeclarePairedDelimiter\bra{\langle}{\rvert}				
\DeclarePairedDelimiter\ket{\lvert}{\rangle}				
\DeclarePairedDelimiterX\braket[2]{\langle}{\rangle}{#1 \delimsize\vert #2}			
\DeclarePairedDelimiterX\ketbra[2]{\vert}{\vert}{#1 \rangle\delimsize\langle #2}	
\DeclarePairedDelimiter\prom{\langle}{\rangle}				

\DeclarePairedDelimiter\Bparen{\Big(}{\Big)}
\DeclarePairedDelimiter\bparen{\big(}{\big)}

\DeclarePairedDelimiter\pen{\llbracket}{\rrbracket}				

\DeclarePairedDelimiterX\AConm[2]{\big\lbrace}{\big\rbrace}{#1 \,\text{\LARGE,}\, #2}		
\DeclarePairedDelimiterX\Conmu[2]{\big[}{\big]}{#1 \,\text{\LARGE,}\, #2}					

\DeclareMathOperator{\sgn}{sgn}
\DeclareMathOperator{\Si}{Si}

\makeatletter
\def\@email#1#2{%
 \endgroup
 \patchcmd{\titleblock@produce}
  {\frontmatter@RRAPformat}
  {\frontmatter@RRAPformat{\produce@RRAP{*#1\href{mailto:#2}{#2}}}\frontmatter@RRAPformat}
  {}{}
}%
\makeatother
\begin{document}

\title[]{Exact Green's functions for localized irreversible potentials}
\author{J. I. Castro--Alatorre}
	\affiliation{%
	Instituto de F\'isica, Benem\'erita Universidad Aut\'onoma de Puebla, Apartado Postal J-48, 72570 Puebla, M\'exico
}%
\author{D. Condado}
	\affiliation{%
	Instituto de F\'isica, Benem\'erita Universidad Aut\'onoma de Puebla, Apartado Postal J-48, 72570 Puebla, M\'exico
}%
\author{E. Sadurn\'i}%
	\email{emerson.sadurni@gmail.com}
	\affiliation{%
	Instituto de F\'isica, Benem\'erita Universidad Aut\'onoma de Puebla, Apartado Postal J-48, 72570 Puebla, M\'exico
}%

\date{\today}

\begin{abstract}
	We study the quantum-mechanical problem of scattering caused by a localized obstacle that breaks
	spatial and temporal reversibility. Accordingly, we follow Maxwell's prescription to achieve a violation of
	the second law of thermodynamics by means of a momentum-dependent interaction in the Hamiltonian,
	resulting in what is known as Maxwell's demon.
	We obtain the energy-dependent Green's function analytically,
	as well as its meromorphic structure. The poles lead directly to the solution of the evolution problem, in the spirit of M. Moshinsky's work in the 1950s. Symmetric initial conditions are evolved in this way,
	showing important differences between classical and wave-like irreversibility 
	in terms of collapses and revivals of wave packets.
	Our setting can be generalized to other wave operators, e.g. electromagnetic cavities in a classical regime.
\end{abstract}

\maketitle
\newpage


\section{\label{sec:level1}Introduction}

Explicit time-dependent solutions of the Schr\"odinger equation have been of interest since the advent of quantum mechanics. In 1952, M. Moshinsky showed that the evolution of a particle beam emerging from a shutter displays important details inherent to interference effects. In the same decade, Moshinsky linked this study to the poles of the S matrix in scattering theory, including irreversible problems e.g. resonances and nuclear decay. 

	In this article, we study a closed dynamical system with spatial and temporal irreversibility, using similar techniques but in a modern context: The microscopic test of the second law of thermodynamics. Our closed system consists of an arbitrarily large ensemble of independent particles described by the Sch\"odinger equation under the influence of a momentum-dependent potential localized in some small region -- an obstacle. 
	In the classical regime, the so-called Maxwell's demon \cite{Maxwell1871Theory} falls into this class of problems,
	whereby the process of discriminating particles by their velocity is called Maxwellian irreversibility.
	With this in mind,
	it is possible to study the quantum effects of an interaction potential that depends on the momentum of a wave, i.e. an operator that addresses its Fourier component.
	Such momentum-dependent interactions have been used extensively to model nuclear,
	molecular or even relativistic dynamics \cite{Dorso1993,Cordero1995,Nara2020}.
	Thus, we expect that a point-like defect that operates on a particle according to its velocity,
	should reproduce reasonably well the classical division of fast and slow components of an ensemble into two compartments,
	plus interference effects that we shall discuss carefully in our treatment.
	
	The mathematical goal of this work is to obtain in closed form
	the corresponding energy-dependent Green's function,
	thus providing an analytical solution to scattering and time-dependent evolution via Laplace inversion.
	It should be noted that such a function will have broken exchange symmetry (e.g., $G(x,x')\neq G(x',x)$),
	as is to be expected for a system with Maxwellian irreversibility or broken time reversal invariance.
	Explicit Green's functions with such properties are scarce in the literature \cite{GroscheSteiner1998, Schulman2005},
	so our results shall include new formulae for these objects.
	
	In terms of applications,
	the limitations of the second-law of thermodynamics have been discussed since the appearance of Maxwell's demon\cite{
	Szilard1929, Brillouin1951, Landauer1961, Bennett1982, Lloyd1997, Maruyama2009, Plesch2014, LeffRex2014},
	but they have not been emulated dynamically so far in the quantum realm;
	instead, informational treatments have been used which, in general, are based on measurements and feedback
	\cite{Scully2001,PriceBannerman2008,Camati2016,Cottet2017,Elouard2017} in various types of arrangements,
	such as photonic setups \cite{Ruschhaupt2006,Vidrighin2016}, ultracold atoms \cite{Kumar2018},
	superconducting quantum circuits \cite{Masuyama2018}, QED cavities \cite{NajeraSantos2020},
	quantum dots \cite{AnnbyAndersson2020} and electronic circuits \cite{Chida2017,Schaller2018}.
	In view of this,
	our approach shall be ideal for applications that involve wave dynamics of a broader type,
	without wave collapse mechanisms; electromagnetic cavities can be considered if one perturbs the Helmholtz operator with complex terms, as in dielectric media.

	In section \ref{sec:MIP} we introduce a momentum-dependent potential in a classical Hamiltonian
	that exerts Maxwellian irreversibility on the particles involved,
	and then we generalize it to the quantum mechanical domain.
	Subsequently, the formalism of irreversible non-symmetric Green's functions is introduced, obtaining thereby a new closed expression.
	Lastly, in section \ref{sec:APC} we present a dynamical analysis of symmetric initial conditions.
	
\section{Maxwellian Irreversible Problems with Localized Interactions}\label{sec:MIP}
\begin{figure}[b]\centering
		\includegraphics[width=0.45\textwidth]{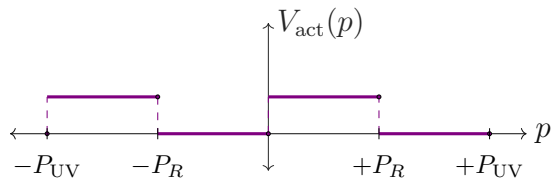}
		\caption{Activation potential $V_\text{act}(p)$ defined in \eqref{eq:VactSgn}
			with reference momentum $P_R$ and ultraviolet cleaving $P_\text{UV}$.}
		\label{fig:VactPUV}
	\end{figure}

	We motivate our discussion with a classical problem that consists of
	an ensemble of particles in an origin-centered container with a length of $2x_L$.
	In this context, particles are considered independent,
	therefore the action of a potential is separable
	and a Hamiltonian formulation per particle is possible.
	For this system, the Hamiltonian takes the form
	\begin{equation}\label{eq:HamV}
		H= \frac{p^2}{2m} + V_\text{box}(x) + V_\text{int}(x,p),
	\end{equation}
	where $V_\text{box}$ represents a pair of impenetrable barriers at $x=\pm x_L$, $V_\text {int}$ is a momentum-dependent potential with strength $V_0$ at $x =0$, i.e.,
	\begin{equation}\label{eq:Vcl}
		V_\text{int}(x,p) = V_0 \delta(x)V_\text{act}(p),
	\end{equation}
	and $V_\text{act}(p)$ is an activation function that determines whether particles remain on one side of the container or pass to the other side according to the particle's momentum.
	In particular, we have the following expression
	\begin{equation}\label{eq:VactSgn}\begin{split}
		V_\text{act}(p) &= f_-\bparen{|p|} \sgn(p) + f_+\bparen{|p|}
	\end{split}\end{equation}
	with
	\begin{align}\begin{autobreak}
		2 f_\pm\bparen{|p|} = \Theta\Bparen{P_R-|p|} \pm \Theta\Bparen{|p|-P_R} \mp \Theta\Bparen{|p|-P_\text{UV}}
	\end{autobreak}\end{align}
	where the reference momentum $P_R$ will discern which particles will be influenced by the perturbation,
	depending on their momentum, whereas the $\delta$ distribution is a contact interaction depending on particle's position.
	In addition, an ultraviolet cutoff $P_\text{UV}$ can be introduced in case the potential does not operate at high frequencies; for instance, in electromagnetic realizations $P_\text{UV}$ is necessary, as dielectric materials operate in specific ranges (see fig.~\ref{fig:VactPUV}).
	
\begin{figure}[h]\centering
		\includegraphics[width=0.45\textwidth]{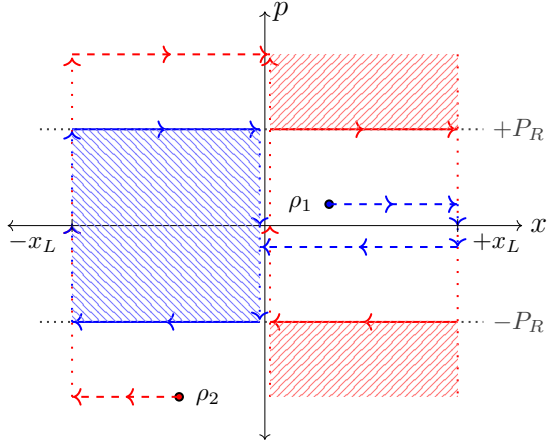}
		\caption{Phase space evolution of two particle ensembles $\rho_1(|p|<|P_R|)$ \& $\rho_2(|p|>|P_R|)$.
		The activation potential separates particles according to the reference momentum, leaving two well differentiated zones.}
		\label{fig:EFEnsamble}
	\end{figure}

	To appreciate the effect of this potential, suppose two ensembles $\rho_1$ and $\rho_2$, shown in fig.~\ref{fig:EFEnsamble}:
	$\rho_1$ represents a collection of independent particles in the first quadrant with a right-directed momentum less than $P_R$,
	therefore, when the system evolves, the phase space corresponding to the zone $ -x_L <x <0 $ and $ | p | <| P_R | $ will be filled.
	Conversely, $ \rho_2 $ represents a collection of independent particles in the third quadrant with a left-directed momentum greater than $ P_R $,
	so,	when the system evolves, the phase space corresponding to the zone $ 0 <x <x_L $ and $ | p |> | P_R | $ will be filled.
	It should also be noted that, after the selection process has taken place,
	the system reaches an equilibrium where each compartment possesses temperatures $T_{1,2}$ such that
	\begin{align}
		T_2\Bparen{x>0} > T_1\Bparen{x<0}.
	\end{align}
	Therefore $V_\text{act}(p)$ effectively separates the particles into two well differentiated zones according to their momentum.	
	

	\subsection{Quantum Mechanical Generalization} \label{sec:QMG}
	Now we can omit the presence of $V_\text{box}$ in the Hamiltonian operator by introducing Dirichlet boundaries. It is also important to preserve the hermiticity of $H$.
	To this end, we first promote $V_\text{int}(x,p)$ in \eqref{eq:Vcl} to an operator
	\begin{align}\label{eq:PotHer}\begin{autobreak}
		\hat{V}_\text{int}(\hatx,\hatp) \equiv V_0 \delta(\hatx) \hat{V}_\text{act}(\hatp),
	\end{autobreak}\end{align}
	which must be symmetrized in order to get a hermitian potential.
	Note that the action of $\hat{V}_\text{act}(\hatp)$ on a complete basis of plane waves is
	$\hat{V}_\text{act}(\hatp)\exp(ikx) = V_\text{act}(\hbar k)\exp(ikx)$,
	where $V_\text{act}(\hbar k)$ is just a number as specified by the classical function \eqref{eq:VactSgn}. Therefore, the action of (\ref{eq:PotHer}) on any wave given by $\psi(x)=\langle x|\psi \rangle$ can be defined by means of the action of $\hat{V}_\text{act}(\hatp)$ on the state $\ket{\psi}$; we have
\begin{align} \label{actonpsi} \begin{split}
 \hat{V}_\text{act}(-i\hbar\nabla)\langle x | \psi \rangle& = \bra{x} \hat{V}_\text{act}(\hatp) \ket{\psi} = \int_{-\infty}^{\infty}dp \bra{x} \hat{V}_\text{act}(\hatp) \ket{p}\langle p|\psi \rangle = \int_{-\infty}^{\infty}dp V_\text{act}(p) \langle x|p \rangle\langle p|\psi \rangle \\ 
 & =\int_{-\infty}^{\infty}dp \int_{-\infty}^{\infty}dx' V_\text{act}(p) \langle x|p \rangle\langle p|x' \rangle \langle x'|\psi \rangle
=\int_{-\infty}^{\infty}dx' \int_{-\infty}^{\infty}dp V_\text{act}(p) \frac{e^{ip(x-x')/\hbar}}{2\pi\hbar} \langle x'|\psi \rangle \\
& = \int_{-\infty}^{\infty}dx' \widetilde{V}_\text{act}(x'-x) \psi(x'),
\end{split}\end{align}  
	where $\widetilde{V}_\text{act}$ is the Fourier transform of $V_\text{act}$. The symmetrization of the operator (\ref{eq:PotHer}) is the following hermitian potential $\hat{V}=\tfrac{1}{2}(\hat{V}_\text{int}+\hat{V}_\text{int}^\dagger)$, and it renders the Schr\"odinger equation as
	\begin{align}\label{eq:DdMQMEq}\begin{autobreak}
		-\frac{\hbar^2}{2m}\nabla^2\Psi(x,t)
		+ \frac{V_0}{2}\left[ \delta(\hatx)\hat{V}(-i\hbar\nabla) + \hat{V}(-i\hbar\nabla) \delta(\hatx)\right]\Psi(x,t)
		= i\hbar\partial_t\Psi(x,t).
	\end{autobreak}\end{align}
	The goal is to solve \eqref{eq:DdMQMEq} using energy-dependent Green's functions.
	We emphasize that eigenfunctions are not necessary to obtain the spectral decomposition in closed form.
	An example of this can be seen in appendix \ref{ap:a},
	where the Green's function for a $\delta(x)$-potential is calculated solely using the integrals in the Lippmann-Schwinger equation.
	
	\subsection{A Theorem on Non-Symmetric Green's Functions} \label{sec:NSGf}
	It is known that Green's functions are not always symmetric:
	the cases in which symmetry under exchange of spatial variables is recovered correspond to real Hamiltonians and time-reversibility.
	To see this, we present the following elementary theorem:
	
	\begin{theorem*}
	Let $\hat G^{(\pm)}$ be a solution of
	$\Bparen{\hat H-E}\hat G^{(\pm)} = \mathbb{I}$
	and
	$\hat G^{(\pm)}\Bparen{\hat H-E} = \mathbb{I}$,
	where $\hat H$ is Hermitian and $\hat G^{(\pm)}$ in the position-basis is
	$\bra{x}\hat{G}^{(\pm)}\ket{x'}
		= \lim_{\varepsilon\rightarrow0^+} \bra{x}\frac{1}{\hat H-E\mp i\varepsilon}\ket{x'}
		= G^{(\pm)}(x,x';E)$.
	Then $\Bparen{ G^{(\pm)}(x,x';E) }^* = G^{(\mp)}(x',x;E))$.
	\end{theorem*}
	
	\begin{proof*}
	In the position-basis, $\hat G^{(\pm)}$ must fulfill
	\begin{subequations}
	\begin{equation}
		\Bparen{H(x,-i\partial_x)-E} G^{(\pm)}(x,x';E) = \delta(x-x')
	\end{equation}
	and
	\begin{equation}
		\Bparen{H^*(x',-i\partial_{x'})-E} G^{(\pm)}(x,x';E) = \delta(x-x').
	\end{equation}
	\end{subequations}
	
	Taking a Hamiltonian such that $H=H^\dagger$, it follows that
	\begin{subequations}
	\begin{equation}\label{eq:pr1}
		\hat G^{(\pm)^\dagger} = \hat G^{(\mp)},
	\end{equation}
	or, expressed in the position basis
	\begin{equation}
		\bra{x} \hat G^{(\pm)^\dagger} \ket{x'} = \bra{x} \hat G^{(\mp)} \ket{x'}.
	\end{equation}
	\end{subequations}
	Note that the left-hand side becomes
	\begin{equation*}
		\bra{x} \hat G^{(\pm)^\dagger} \ket{x'} = \bparen{ \bra{x'}\hat G^{(\pm)}\ket{x} }^*
	\end{equation*}
	after taking out the conjugate transpose.
	Consequently,
	\begin{equation}
		\Bparen{ G^{(\pm)}(x',x;E) }^* = G^{(\mp)}(x,x';E).
	\end{equation}
	So the advanced and retarded Green's function, for this case,
	are related to an index exchange and complex conjugation. $\blacksquare$
	\end{proof*}
	
	\begin{corollary*}
		$G^{(\pm)}(x,x';E)$ is symmetric if and only if the Hamiltonian is real.
	\end{corollary*}
	
	\begin{proof*}
	$\Rightarrow)$
	By taking a Hamiltonian such that $H=H^*$, it follows that
	\begin{subequations}
	\begin{equation}
		\hat G^{(\pm)^*} = \hat G^{(\mp)}
	\end{equation}
	or, expressed in the position basis
	\begin{equation}
		\bra{x} \hat G^{(\pm)^*}\ket{x'} = \bra{x} \hat G^{(\mp)} \ket{x'}.
	\end{equation}
	\end{subequations}
	Note that the left-hand side can be transformed into
	\begin{equation*}
		\bra{x} ((\hat G^{(\pm)^*})^\text{T})^\text{T} \ket{x'}
		= \bra{x} (\hat G^{(\pm)^\dagger})^\text{T} \ket{x'}
		= \bra{x} (\hat G^{(\mp)})^\text{T} \ket{x'},
	\end{equation*}
	where \eqref{eq:pr1} was used in the last step.
	Consequently,	
	\begin{equation}
		\bra{x'} (\hat G^{(\mp)}) \ket{x} = \bra{x} \hat G^{(\mp)} \ket{x'}.
	\end{equation}
	Therefore, $G^{(\pm)}(x,x';E)$ is symmetric.
	
	$\Leftarrow)$
	Suppose that
	\begin{subequations}
	\begin{equation}
		G^{(\pm)}(x,x';E) = G^{(\pm)}(x',x;E),
	\end{equation}
	or, expressed in Dirac notation
	\begin{equation}
		\bra{x} \hat G^{(\pm)} \ket{x'} = \bra{x'} \hat G^{(\pm)} \ket{x}.
	\end{equation}
	\end{subequations}
	Note that the right-hand side can be transformed into
	\begin{equation*}
		\bra{x'} (\hat G^{(\pm)^\dagger})^\dagger \ket{x}
		= \bra{x'} \hat G^{(\mp)^\dagger} \ket{x}
		= \bra{x} \hat G^{(\mp)^*} \ket{x'}
	\end{equation*}
	where \eqref{eq:pr1} was used in the middle step.
	Consequently,	
	\begin{equation}
		\bra{x} \hat G^{(\pm)} \ket{x'} = \bra{x} \hat G^{(\mp)^*} \ket{x'}.
	\end{equation}	
	Therefore, $\hat G^{(\pm)^*} = \hat G^{(\mp)}$, and the Hamiltonian is real.
	$\blacksquare$
	\end{proof*}

	\subsection{Exact Form of Green's Function}	\label{sec:EFGf}
	In this work we focus on the analysis of a function $ G_\text{Demon} $ that solves the following problem
	\begin{align}\label{eq:GreenDem}
		\Bparen{H+V(x,p)-E}G_\text{Demon}= \delta(x-x'),
	\end{align}
	where $H$ is any Hamiltonian whose Green's function $G_0$ is known and $V(x,p)$ is the Maxwellian interaction in \eqref{eq:DdMQMEq}. From here on we work with units $\hbar=1$.
	We recall that a particle with a Dirac-delta potential gives rise to an equation with a source,
	similar to the Lippmann-Schwinger equation.
	This allows an exact solution for the energy-dependent Green's function
	by an evaluation of the corresponding integrals in the first term of the Born series.
	Note however that when the potential is affected by a momentum-dependent activation function (hence irreversible),
	the integral is more involved, as $G$ is self-contained in the expressions. For this reason,
	we must solve an integral equation as well as a functional relation.	
	The explicit equation in operator-form to be solved is
	\begin{equation}\label{eq:GreenVxp}
		\Bparen{\hat H-E +V(\hatp)\delta(\hatx)+\delta(\hatx)V(\hatp)} \hat G_p = \mathbb{I}.
	\end{equation}
	Inspired by the method employed for solving the problem with a delta perturbation
	that depends only on the position (see appendix \ref{ap:a}),
	the following integral equation is obtained
	\begin{align}\label{eq:GreenPInt}\begin{split}
		G_p(x,x',E)& = G_0(x,x',E) \\
			&- \int dy\,G_0(x,y,E)\delta(y)V(\hatp)G_p(y,x',E) \\
			&- \int dy\,G_0(x,y,E)V(\hatp)\delta(y)G_p(y,x',E).				
	\end{split}\end{align}
	where the momentum operator $\hat p$ in the expression above is understood as $-i\partial_y$.
	Prior to the evaluation of (\ref{eq:GreenPInt})
	we insert another complete set in each integral in the form
	\begin{subequations}\label{eq:GreenPInt2}
		\begin{align}\begin{split}
			& \bra{y}\delta(\hat x)V(\hatp)\hat G_p \ket{x'}
				= \int dy'\,\bra{y}\delta(\hat x)V(\hatp)\ketbra{y'}{y'} \hat G_p \ket{x'} \\
				&= \int dy'\,\frac{\delta(y)}{2\pi}\int dp\,e^{ip(y-y')}V(p)G_p(y',x',E) \\
				&= \delta(y)\int dy'\,\widetilde V(y')G_p(y',x',E),
		\end{split}\end{align}
		where a complete set of plane waves was introduced in the second line, and as before
		\begin{equation}
			\widetilde V(y') = \frac{1}{2\pi}\int dp\, e^{-ipy'}V(p),
		\end{equation}
		is the Fourier transform of the potential;
		whereas, for the second integral, we have
		\begin{align}\begin{split}
			& \bra{y}V(\hatp)\delta(\hat x)\hat G_p \ket{x'}
				= \int dy'\,\bra{y}V(\hatp)\delta(\hat x)\ketbra{y'}{y'} \hat G_p \ket{x'} \\
				&= \int dy'\,\frac{\delta(y')}{2\pi}\int dp\,e^{ip(y-y')}V(p)G_p(y',x',E) \\
				&= \widetilde V(-y)\int dy'\,\delta(y')G_p(y',x',E) \\
				&= \widetilde V(-y) G_p(0,x',E).
		\end{split}\end{align}
	\end{subequations}
	Substitution of \eqref{eq:GreenPInt2} in \eqref{eq:GreenPInt}, leads to
	\begin{align}\begin{split}
		&G_p(x,x',E) = G_0(x,x',E)- G_0(x,0,E)\int dy'\,\widetilde V(y')G_p(y',x',E) 
			- \int dy\,G_0(x,y,E)\widetilde V(-y) G_p(0,x',E).
	\end{split}\end{align}
	In order to get $G_p(x,x',E)$, we first multiply the last expression by $\widetilde V(x)$ and integrate over $x$,
	\begin{align}\begin{split}
		&\int dx\,\widetilde V(x)G_p(x,x',E) = \quad \int dx\,\widetilde V(x)G_0(x,x',E)
			- \int dx\,\widetilde V(x)G_0(x,0,E)\int dy'\,\widetilde V(y')G_p(y',x',E) \\
			&- \int dx\,\widetilde V(x) \int dy\,G_0(x,y,E)\widetilde V(-y) G_p(0,x',E),
	\end{split}\end{align}
	and recognizing that the integral on the left-hand side is the same as the one in the 2nd term of the right-hand side
	(with another integration variable), a consistency condition is obtained:
	\begin{equation}
		\int dy'\,\widetilde V(y')G_p(y',x',E) = \frac{P_1(x',E)-Q_1(E)G_p(0,x',E)}{1+Q_2(E)},
	\end{equation}
	where
	\begin{subequations}
	\begin{equation}
		P_1(x',E) = \int dx\,\widetilde V(x)G_0(x,x',E),
	\end{equation}
	\begin{equation}
		P_2(x,E) = \int dy\,G_0(x,y,E)\widetilde V(-y),
	\end{equation}
	\begin{equation}
		Q_1(E) = \int dx\,\int dy\,\widetilde V(x)G_0(x,y,E)\widetilde V(-y),
	\end{equation}
	\begin{equation}
		Q_2(E) = \int dx\,\widetilde V(x)G_0(x,0,E),
	\end{equation}
	\end{subequations}	
	Substituting the above equation again into $G_p(x,x',E)$ given by \eqref{eq:GreenPInt},
	leads to the following functional equation
	\begin{align}\begin{split}
	&G_p(x,x',E) = G_0(x,x',E) - P_2(x,E)G_p(0,x',E)  \\ 
           & - G_0(x,0,E)\frac{P_1(x',E)-Q_1(E)G_p(0,x',E)}{1+Q_2(E)}.
	\end{split}\end{align}
	The latter is not yet a closed formula for $G_p$, for it depends on $G_p$ again.
	Evaluating at $x=0$ provides the reduced functional equation
	\begin{align}\begin{split}
		&G_p(0,x',E) = G_0(0,x',E) - P_2(0,E)G_p(0,x',E) \\
			&- G_0(0,0,E)\frac{P_1(x',E)-Q_1(E)G_p(0,x',E)}{1+Q_2(E)},
	\end{split}\end{align}
	which can be solved for $G_p(0,x',E)$, leaving
	\begin{subequations}
		\begin{equation}
			G_p(0,x',E) = \frac{G_0(0,x',E)-G_0(0,0,E)R_1(x',E)}{1+P_2(0,E)-G_0(0,0,E)Q_3(E)},
		\end{equation}
		where
		\begin{align}
			R_1(x',E) = \frac{P_1(x',E)}{1+Q_2(E)}, && Q_3(E) = \frac{Q_1(E)}{1+Q_2(E)}.
		\end{align}
	\end{subequations}
	The last equation finally is substituted into $G_p(x,x',E)$,
	to obtain its final expression in terms of $G_0(x,x',E)$
	\begin{align}\label{eq:GreenPotencialDelta}\begin{split}
G_p(x, &x', E) = G_0(x,x',E)
+\frac{G_0(x,0,E)G_0(0,x',E)Q_3(E)}{1+P_2(0,E)-G_0(0,0,E)Q_3(E)} \\
& - \frac{G_0(x,0,E)R_1(x',E)\Bparen{1+P_2(0,E)} }{1+P_2(0,E)-G_0(0,0,E)Q_3(E)}
 -\frac{P_2(x,E)\Bparen{G_0(0,x',E)-G_0(0,0,E)R_1(x',E)} }{1+P_2(0,E)-G_0(0,0,E)Q_3(E)}.
\end{split}\end{align}
	We shall see that \eqref{eq:GreenPotencialDelta} can be split into symmetric and antisymmetric contributions,
	where the latter are associated with irreversibility,
	as we have seen from the theorem in the previous section.
	
\section{Application to a particle in a container}\label{sec:APC}
	We now specialize in the case where particles are in a container.
	For this purpose, we employ Dirichlet boundary conditions at the walls of a one-dimensional box.
	The Green's function and the energies are well known for the unperturbed problem.
	Our plan is as follows: first we apply our new Green's function formula to the case of the container with an irreversible perturbation inside; we
	give the explicit form of its spectral decomposition, and we analyze its meromorphic structure in order to find its poles.
	Subsequently, we focus on the evolution problem; therefore,
	we shall need an appropriate definition of entropy that accounts for the emergence of disorder in energy space,
	so a basis-dependent entropy is suggested.
	The next subsection is devoted to the use of Shannon's entropy in our evolution problem.
	Afterwards, we address the explicit problem of numerical evolution by means of spectral decomposition in a discretized space.
	Efficient numerical evaluations are best achieved if this discretization is restricted to a region where the dispersion relation is well approximated by a parabola.
	Therefore, we include a careful analysis of the dispersion relation in the spatially-discretized version of the problem.
	Lastly, we construct specific initial conditions that are completely symmetric and analyze how the wave packet propagates inside the container asymmetrically.
	The reason is obviously the inherent broken spatial symmetry of the problem,
	i.e., the transformation $x\rightarrow -x$, $p\rightarrow -p$ changes the form of the Hamiltonian,
	as both parity and time reversal invariance are absent.
	Then we add a special definition of temperature (or effective beta parameter).
	Therewith we can analyze other types of time-evolving distributions.
	In this part, it is important to show how the entropy can indeed decrease as a function of time,
	resulting in a special kind of ordering or sorting of fast and slow particles, produced by the non-reversible Maxwellian potential.
	
	We start with a free Green's function in a container $G_0^C$, i.e.,
	\begin{align}\label{g1}\begin{autobreak}
		G_0^\text{C}(x,x',E) =
		\frac{2}{L}\sum_{m=1}\frac{\sin(\kappa_{2m}x)\sin(\kappa_{2m}x')}{E_{2m}-E}
			+ \frac{2}{L}\sum_{m=1}\frac{\cos(\kappa_{2m-1}x)\cos(\kappa_{2m-1}x')}{E_{2m-1}-E}
	\end{autobreak}\end{align}
	with $\kappa_n=n\pi/L$ and eigenenergies $E_n = \frac{1}{2}\hbar^2\kappa_n^2$. Although this problem does not have asymptotic states, the perturbed system can be regarded as a scattering problem for waves inside the box. It is important to clarify that our approach using plane waves in (\ref{actonpsi}) is still valid here, if we recognize that the space of functions within the container $L_2(\left[ -L,L \right])$ coincides with the space of functions in $L_2(\mathbf{R})$ whose compact support is restricted to the interval. Their Fourier transform is well defined $\tilde f(p) = \int_{-\infty}^{+\infty}dx e^{ipx} f(x) = \int_{-L}^{+L}dx e^{ipx} f(x)$, and all previous formulae apply here. Conveniently, a basis of plane waves also defines the action of $\hat{V}_\text{act}(\hatp)$ on the space of functions of compact suppport. One may as well resort to a basis of box functions for this purpose, but computations would be more involved. 
Now let us employ our new result: The Green's function (\ref{g1}) can be put in terms of Jacobi's theta function as reported by \citet{GroscheSteiner1998}.
	The Green's function in \eqref{eq:GreenPotencialDelta} with the container is
	\begin{align}\begin{split}
G_p^\text{C}(x,x',E) & =  G_0^\text{C}(x,x',E)
+\frac{P_1^\text{C}(x,E)G_0^\text{C}(0,x',E)-G_0^\text{C}(x,0,E)P_1^\text{C}(x',E)}{1-G_0^\text{C}(0,0,E)Q_1^\text{C}(E)} \\
&+\frac{G_0^\text{C}(x,0,E)G_0^\text{C}(0,x',E)Q_1^\text{C}(E)}{1-G_0^\text{C}(0,0,E)Q_1^\text{C}(E)}
-\frac{P_1^\text{C}(x,E)G_0^\text{C}(0,0,E)P_1^\text{C}(x',E)}{1-G_0^\text{C}(0,0,E)Q_1^\text{C}(E)},
	\end{split}\end{align}
	where $P_1^\text{C}$ and $Q_1^\text{C}$ are the integrals evaluated with $G_0^C$.
	From the previous expression the identification of the symmetric and antisymmetric part of the Green's function is effortless.
	Note that terms that contribute to the antisymmetric part come from the Maxwellian perturbation,
	as the first and third terms are manifestly symmetric.

	\subsection{Pole Structure Analysis}
		The integrals in \eqref{eq:GreenPotencialDelta} can be done in terms of sine-integral functions\cite{Abramowitz1965} (Si)
		using in addition the Fourier's transform of the potential described in fig.~\ref{fig:VactPUV}, e.g.
		\begin{subequations}
		\begin{equation}
			\widetilde V(\pm y) = \frac{\pm 1}{2i\pi y}\Bparen{1-2\cos P_R y},
		\end{equation}
		\begin{align}\begin{split}
			&P_1^\text{C}(x',E) = - P_2^\text{C}(x',E) = \\ &-\frac{2}{i\pi L}
			\sum_{n=1}\frac{\Si\bparen{\xi_+}-\Si\bparen{\xi_-}-\Si(n\pi)}{E_{2n}-E}\sin(\kappa_{2n}x'),
		\end{split}\end{align}
		\begin{equation}
			Q_1^\text{C}(E) =
			\frac{2}{\pi^2L}\sum_{n=1} \frac{\Bparen{\Si\bparen{\xi_+}-\Si\bparen{\xi_-}-\Si(n\pi) }^2}{E_{2n}-E},
		\end{equation}
		\begin{equation}
			Q_2^\text{C}(E) = 0,
		\end{equation}
		with
		\begin{equation}
			\xi_\pm = \bparen{P_R\pm\kappa_{2n}}\frac{L}{2}.
		\end{equation}
		\end{subequations}
		Given the behavior of the $ \Si $ function in $ P_1^\text{C}$ and $Q_1^\text{C}(E)$, the following approximation can be made.
		The argument is written as
		\begin{align}\begin{split}
			&\Si\bparen{n\pi+a} - \Si\bparen{n\pi-a} \simeq \\
			&\frac{\pi}{2}-\frac{\pi}{2}\Theta\Bparen{n-\pen{a/\pi}}
			+\frac{\pi}{2}\Theta\Bparen{\pen{a/\pi}-n}+\pi\epsilon\,\delta_{n,\pen{a/\pi}}.
		\end{split}\end{align}
		where $a= P_R L/2$, $ \pen {a/\pi} $ represents the integer part and
		$\epsilon$ the fractional part of $a/\pi$
		and the step function is zero for the case $n=\pen{a/\pi}$.
		(The full procedure is in appendix \ref{ap:b}.)
		So the integrals are now approximated by
		\begin{subequations}
		\begin{align}\begin{autobreak}
			P_1^\text{C}(x,E)\simeq -\frac{2}{iL}
			\left(\sum_{n=1}^{\pen{a/\pi}-1}\frac{\sin(\kappa_{2n}x)}{E_{2n}-E}\right. 
			- \frac{1}{2}\sum_{n=1}^\infty\frac{\sin(\kappa_{2n}x)}{E_{2n}-E}
			+ \left.\frac{1}{2}\frac{(1+2\epsilon)}{E_{2\pen{a/\pi}}-E}\sin(\kappa_{2\pen{a/\pi}}x)\right)
		\end{autobreak}\end{align}
		and
		\begin{align}
			Q_1^\text{C}(E)\simeq\frac{1}{2L}\left(\sum_{n=1}^\infty\frac{1}{E_{2n}-E} - \frac{1}{E_{2\pen{a/\pi}}-E}\right).
		\end{align}
		\end{subequations}
		
		The pole structure of the whole antisymmetric term can be very intricate,
		according to the transcendental equation
		\begin{align}\label{eq:GpDen}
			1 - G_0^\text{C}(0,0,E)Q_1^\text{C}(E) = 0,
		\end{align}
		where it is straightforward to see that
		\begin{align*}
			G_0^\text{C}(0,0,E) = \frac{\tan(L\sqrt{2E}/2\hbar)}{\hbar\sqrt{2E}},
		\end{align*}
		\begin{align*}
			Q_1^\text{C}(E) = \frac{1}{2L}\left(
				\frac{1}{2E} - \frac{L\cot(L\sqrt{2E}/2\hbar)}{2\hbar\sqrt{2E}} -\frac{1}{E_{2\pen{a/\pi}}-E} \right),
		\end{align*}
		and clearly $E_{2\pen{a/\pi}}$ is not a zero of \eqref{eq:GpDen},
		since the product $G_0^\text{C}(0,0,E)Q_1^\text{C}(E)$ would tend to infinity if $E\rightarrow E_{2\pen{a/\pi}}$.
		However, it is noted that the numerator
		\begin{align}\label{eq:GpNum}
			P_1^\text{C}(x,E)G_0^\text{C}(0,x',E)-G_0^\text{C}(x,0,E)P_1^\text{C}(x',E)
		\end{align}
		still contains the poles of $G_0^\text{C}$; while the new term $P_1^\text{C}(x,E)$, proportional to
		$\sin(\kappa_{2\pen{a/\pi}}x)/ (E_{2\pen{a/\pi}}-E) $, which in general (if $x\neq x' $) does not disappear,
		contributes to a new pole located at $E_{2\pen{a/\pi}}$.
		To sum up, the antisymmetry under exchange $x\leftrightarrow x'$ in both stationary and
		time-dependent solutions comes predominantly from the harmonic inversion of \eqref{eq:GpNum}
		under the approximation
		\begin{align*}
			\propto \frac{ \sin(\kappa_{2\pen{a/\pi}}x) G_0^\text{C}(x',0,E)
				- G_0^\text{C}(0,x,E) \sin(\kappa_{2\pen{a/\pi}}x') }{E_{2\pen{a/\pi}}-E}
		\end{align*}
		whose pole produces $\exp(-iE_{2\pen{a/\pi}}t/\hbar)$ in the spectral decomposition of the wave function.
			
	\subsection{Shannon's Entropy}
		The entropy is fundamental in the analysis of asymmetric evolution
		inasmuch a dynamic effect of apparent ordering is sought.
		Since the von--Neumann equation without a source does not capture the irreversibility phenomenon,
		a notion of entropy that describes disorder with respect to a specific basis (e.g. energy) is required.
		Shannon's definition of entropy is
		\begin{equation}\label{eq:ShEntropia}
		\sigma_\text{\tiny Sh} = - \sum_m \varrho_m \log \varrho_m.
		\end{equation}
		where the probabilities $\varrho_m$ will be given by the overlap (integral) of the wave function with the basis of the free problem, i.e.
		\begin{equation}
			\varrho_m = \big|\braket{m,E_m^\text{(0)}}{\Psi,t}\big|^2 =
			\left|\sum_{n'}\braket{m,E_m^\text{(0)}}{n'}\braket{n'}{\Psi,t}\right|^2.
		\end{equation}
		Likewise, the total entropy of the system must be estimated separately,
		as Shannon's entropy applies only to the particles inside the container,
		but does not contemplates the reaction of $V(x,p)$ itself, which is not dynamically involved.
		To this end, we estimate the work done by the potential on the trapped wave and {\it vice versa}.
		Also, using the principle of extensivity one can find a lower bound for the total entropy of the system $S_\text{t}$
		as the linear combination of the particle's entropy $S_\text{p}$ plus Maxwellian-potential's entropy $S_\text{d}$, i.e.
		\begin{subequations}
		\begin{equation}
			\Delta S_\text{t} = \Delta S_\text{p} + \Delta S_\text{d} \overset{?}{\geq} 0
		\end{equation}
		with $\Delta S_\text{p}\leq 0$ as the entropy of the particles must decrease because of the Maxwellian potential.
		Furthermore, one can find a lower bound for the Maxwellian-potential's entropy change in terms of the work done by the potential as
		\begin{equation}
			\Delta S_\text{d} = \int\frac{\delta'Q}{T} \;\sim\;
			\frac{\Delta Q}{\prom T} \geq \frac{1}{\prom T}\Delta V = \text{(const.)}\Delta V.
		\end{equation}
		From this, it follows that
		\begin{equation}
			\Delta S_\text{t} \geq \Delta S_\text{p} + \text{(const.)}\Delta V.
		\end{equation}
		\end{subequations}
		Therefore, when taking into account the work done by the Maxwellian potential,
		its contribution must compensate for the partial entropy reduction.
		Indeed, for a container with two separated compartments with volume $\upsilon$,
		the change in the particle's entropy is
		\begin{subequations}
		\begin{equation}
			\Delta S_\text{p} = -\left(\frac{P_R}{T_R} + \frac{P_L}{T_L}\right) \upsilon\log 2,
		\end{equation}
		where $P_{R,L}$ and $T_{R,L}$ are the pressure and temperature for the right (R) and left (L) compartment, computed as ideal gases.
		Moreover, the internal energy $U\propto P\upsilon$, so
		\begin{equation}
			\Delta S_\text{p} \propto - \beta_\text{B}\left( U_R + U_L \right)\log 2,
		\end{equation}
		\end{subequations}
		where $\beta_\text{B}$ is the thermodynamic beta and $U_{R,L}$ the corresponding lateral internal energy.
		
	\subsection{Spatial and Spectral Decomposition}
		We proceed to discretize the Hamiltonian on a lattice.
		This enable us to treat the problem as a matrix representation on a basis of point-like functions.
		Since the treatment is equivalent to a tight binding model in a crystal,
		it is advisable to use the first Brillouin zone to calculate the energies.
		In this way, the activation potential in \eqref{eq:VactSgn} will be non-zero in the intervals $[-\kappa_D,-\kappa_R]$ and $[0,\kappa_R]$,
		resulting in the action zones of the Maxwellian potential according to the reference momentum $P_R\leftrightarrow\kappa_R$.
		This can be seen in the graph below in fig.~\ref{fig:EigenEn}.
		However, it is necessary to work in the quasi-parabolic energy regime that is below the {\it Dirac point} \cite{Yael2016} $(\varpi_\text{D})$,
		obtaining the upper graph, a parabola with regions where the Maxwellian potential acts.
		It is worth mentioning that the region of the potential for $p<-P_R$ is not bounded above as in the graph below.
		\begin{figure}[b]\centering
		\includegraphics[width=0.48\textwidth]{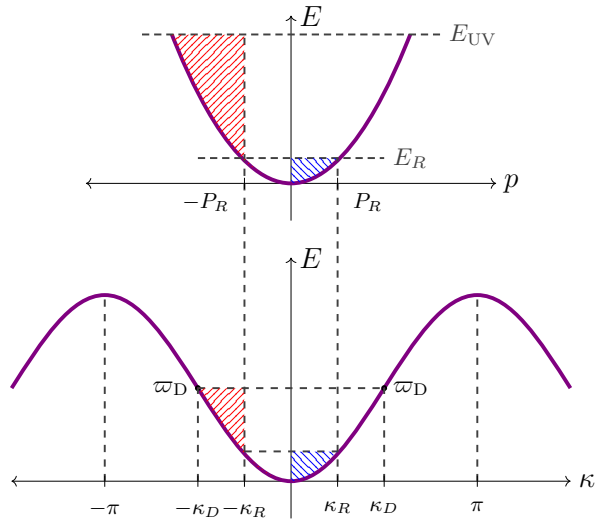}
		\caption{The graph below shows the energies in a tight binding model in a crystal,
		the coloured zones represent the activation potential in \eqref{eq:VactSgn} with reference momentum $P_R\leftrightarrow\kappa_R$.
		Also, using the quasi-parabolic energy regime below the {\it Dirac point} $(\varpi_\text{D})$, the upper graph is obtained.
		}
		\label{fig:EigenEn}
		\end{figure}%
		Therefore, the Hamiltonian's action on a plane wave
		\begin{equation}
		\ket{\kappa} = \frac{1}{\sqrt{2\pi}}\sum_{n} e^{i\kappa n}\ket{n},
		\end{equation}
		is no longer restricted to the unperturbed part plus the defect at the origin,
		instead we have a non-local effect that can be obtained directly by calculating the matrix elements at site $n$
		\begin{align}\begin{split}
		\bra{n}H\ket{\kappa} &= \frac{\hbar^2}{m a^2}\Bparen{1-\cos\kappa}\frac{e^{ikn}}{\sqrt{2\pi}} \\
			&+ \frac{1}{2i\pi n \sqrt{2\pi}}\Bparen{2\cos\bparen{\kappa_R n}-1-e^{-i\kappa_D n}} \\
			&- \sum_{n'}\frac{\delta_{n,0}}{2i\pi n'}\cdot\frac{e^{i\kappa n'}}{\sqrt{2\pi}}
				\Bparen{2\cos\bparen{\kappa_R n'}-1-e^{i\kappa_D n'}}.
	 		\end{split}\end{align}	
	 		where $a$ is the scale parameter.	
		Also, the Hamiltonian will be diagonalized using a discretized basis $\ket{n}$, as many sites as frequencies are necessary, i.e.
		\begin{align}\begin{split}\label{eq:nHnp}
		\bra{n}H\ket{n'} = &- \frac{\hbar^2}{2m a^2}\bparen{\delta_{n-1,n'} - 2\delta_{n,n'} + \delta_{n+1,n'} } \\
			&+ \frac{V_0}{2}\cdot\frac{\delta_{n',0}}{2i\pi n}\Bparen{2\cos\bparen{\kappa_R n}-1-e^{-i\kappa_D n}} \\
			&- \frac{V_0}{2}\cdot\frac{\delta_{n,0}}{2i\pi n'}\Bparen{2\cos\bparen{\kappa_R n'}-1-e^{i\kappa_D n'}}.
		\end{split}\end{align}
		Note that for the central element (the evaluation of the corresponding integrals at $n=0=n'$),
		\begin{align}
			\bra{0}H\ket{0} = \frac{\hbar^2}{2m a^2}\cdot 2 + \frac{V_0}{2}\cdot\frac{\kappa_D}{\pi}.
		\end{align}
		This shows that the potential at its location is finite in a discretized setting, 
		and its intensity $V_0$ can be adjusted at will.
		
		Finally, the wave function $\Psi(t)$ at site $n$ is
		\begin{subequations}
		\begin{equation}
			\braket{n}{\Psi,t} = \sum_m \exp\bparen{-itE_m/\hbar}\braket{n}{m,E_m} \braket{m,E_m}{\Psi_0}
		\end{equation}
		where $E_m$ are the eigenvalues of the problem,
		$\braket{n}{m,E_m}$ are stationary functions i.e. eigenvectors,
		while $\braket{m,E_m}{\Psi_0}$ is the overlap (integral) of the initial condition with the basis, i.e.
		\begin{equation}
		\braket{m,E_m}{\Psi_0} = \sum_{n'} \braket{m,E_m}{n'} \braket{n'}{\Psi_0}.
		\end{equation}
		where $\braket{n'}{\Psi_0}$ is the initial condition.
		\end{subequations}

	\subsection{Dynamical Analysis of Symmetric Initial Conditions}
		Shannon analogue of a Boltzmann thermal distribution \cite{Schleich2001}
		(e.g as understood by superposition of particle's number in photonic states)
		can be used as an appropriate initial condition for box states.
		The idea is to monitor the evolution of such a state and its subsequent ordering. We have:
		\begin{equation}
		\psi_0^\text{B}(\beta,n)= \sum_{q=1}^{N_\text{max}} \exp\Bparen{-\beta\bparen{q^2-1}} \sin\frac{q\bparen{n+N}\pi}{2N},
		\end{equation}
		here $ n $ is the site such that $-N\leq n\leq N$ (the Maxwellian potential is at $n = 0$),
		$N_\text{max}=2N+1$ is the maximum number of $q$ box states that are meaningful in a discretized system,
		and $\beta $ is an order parameter which would correspond to
		\begin{equation}
			\beta = \frac{E_0}{k_\text{B}T}, \quad\text{with}\quad E_0 = \frac{\pi^2\hbar^2}{2mL^2}
		\end{equation}
		in thermodynamics.
		This probability overlaps with the components of the eigenvectors $\nu_m^{(n)}$ of \eqref{eq:nHnp}
		\begin{equation}
		\Psi_0^\text{B}(\beta,m) = \sum_{n=1}^{2N+1} \psi_0^\text{B}(\beta,n) {\nu_m^{(n)}}^*
		\end{equation}
		obtaining the wave function at the rescaled time $\tau$ ($= \hbar t/2ma^2$ [adim])
		\begin{equation}\label{eq:PsiBoltz}
		\Psi_B(\beta,n,\tau) = \sum_{m=1}^{2N+1}\exp\Bparen{-i \tau\,\Xi_m} \Psi_0^\text{B}(\beta,m) \nu_m^{(n)},
		\end{equation}
		where $\Xi$ ($= 2ma^2E/\hbar^2$ [adim]) are the rescaled eigenenergies of \eqref{eq:nHnp}.
					
		\begin{figure}[htpb]\centering
			\includegraphics[width=0.48\textwidth]{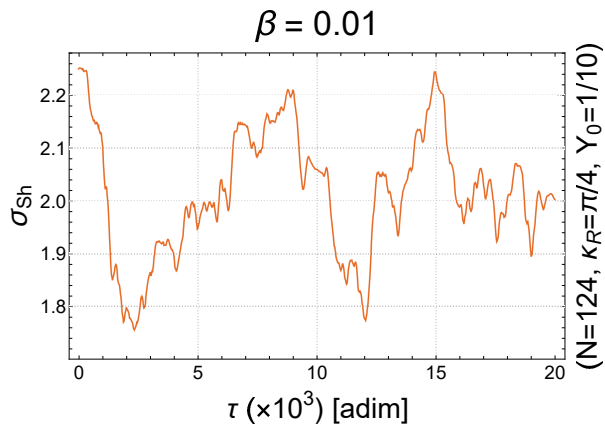}
				\caption{Entropy with $\beta=1/100$.
				A decrease of the entropy is seen at $\tau(\times10^{-3}) \sim$ 2 and 12.}
			\label{fig:EntropyE-SG124-BI4-VG10.png}
		\end{figure}%
		\begin{figure}[htpb]\centering
			\includegraphics[width=0.48\textwidth]{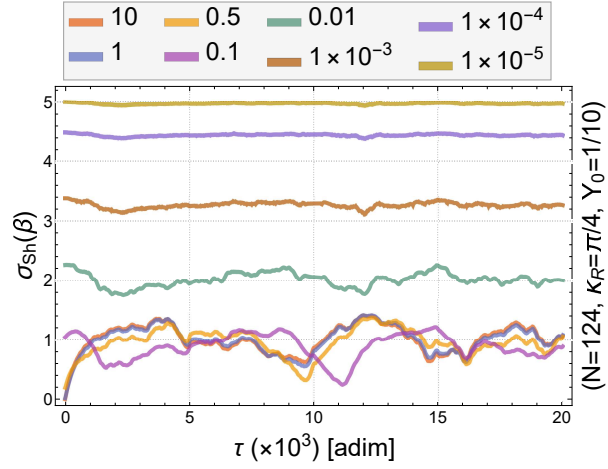}
				\caption{Comparative plot of entropies by varying the temperature value.}
			\label{fig:Entropien-SG124-BI4-VG10.png}
		\end{figure}%
		\begin{figure}[htpb]\centering
			\includegraphics[width=0.48\textwidth]{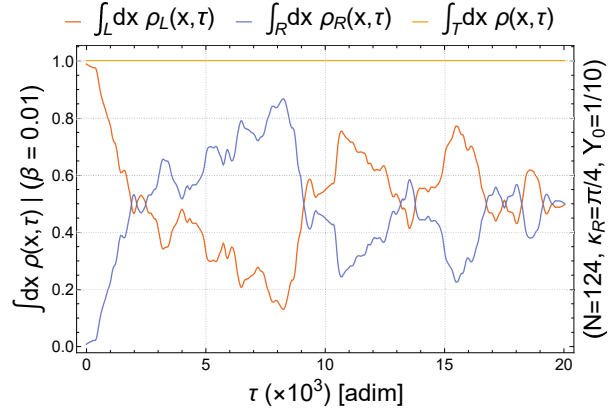}
				\caption{Lateral probabilities for the left (L, red line)
				and right (R, blue line) part of the container with $\beta=1/100$.}
			\label{fig:ProbE-SG124-BI4-VG10.png}
		\end{figure}%
		\begin{figure}[htpb]\centering
			\includegraphics[width=0.48\textwidth]{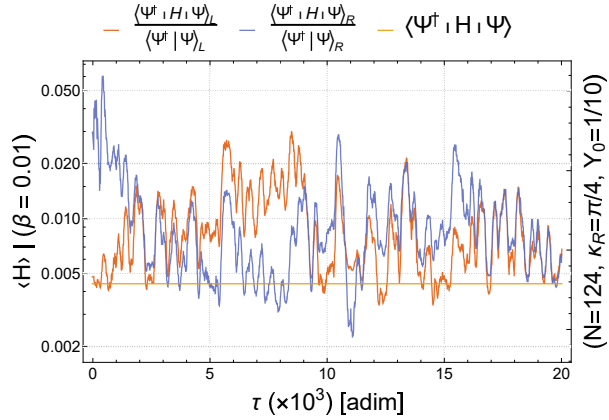}
				\caption{Average internal energy for the left (L, red line)
				and right (R, blue line) part of the container with $\beta=1/100$.}
			\label{fig:PsiHPsiE-SG124-BI4-VG10.png}
		\end{figure}%
		\begin{figure}[htpb]\centering
			\includegraphics[width=0.48\textwidth]{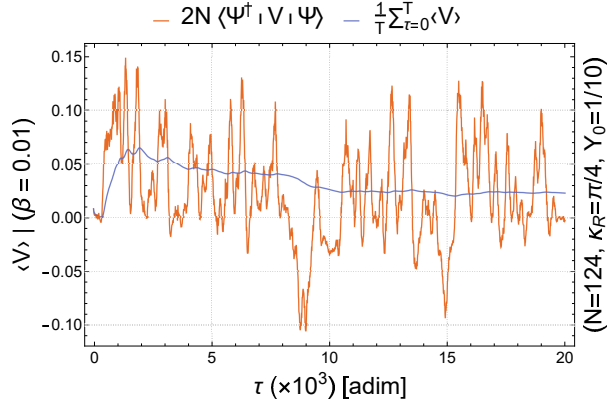}
				\caption{Average Potential Energy. Contributes significantly to the energy balance.
				The blue line indicates the time average at time $\tau$.
				Negative values implies work done by the wave on the Maxwellian potential.}
			\label{fig:PsiVPsiE-SG124-BI4-VG10.png}
		\end{figure}%
		\begin{figure}[htpb]\centering
			\includegraphics[width=0.4\textwidth]{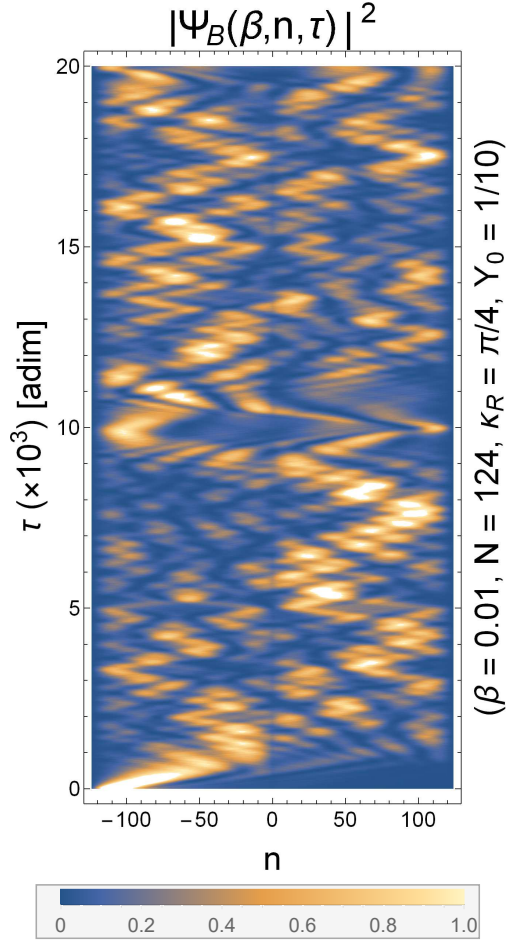}
				\caption{Evolution of a Boltzmann distributed wave packet in the interval
				$-124\leq n\leq 124$ (horizontal axis) at $\tau = 0$ (vertical axis),
				interacting with a Maxwellian potential located at $ n = 0$.
				The colouration exhibits the probability density,
				showing that for $5<\tau (\times10^{-3})< 10$ the wave packet is predominantly on the right side.}
			\label{fig:wBoltzPlotE-SG124-BI4-VG10.png}
		\end{figure}%
		\begin{figure}[htpb]\centering
			\includegraphics[width=0.48\textwidth]{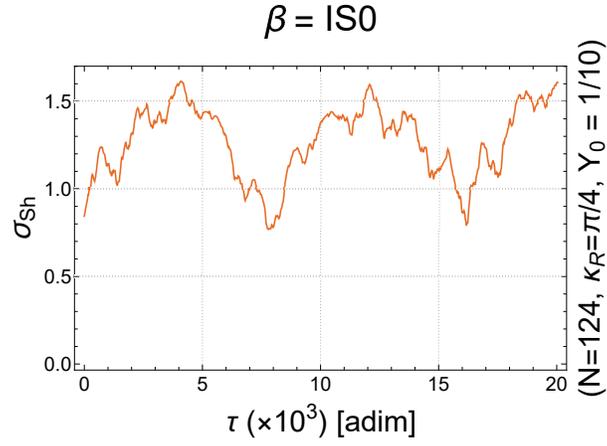}
				\caption{Entropy of a isoespectral wave packet.
				A decrease of the entropy is seen at $\tau(\times10^{-3}) \sim$ 8 and 16.}
			\label{fig:EntropyIS0-SG124-BI4-VG10.png}
		\end{figure}%
		\begin{figure}[htpb]\centering
			\includegraphics[width=0.4\textwidth]{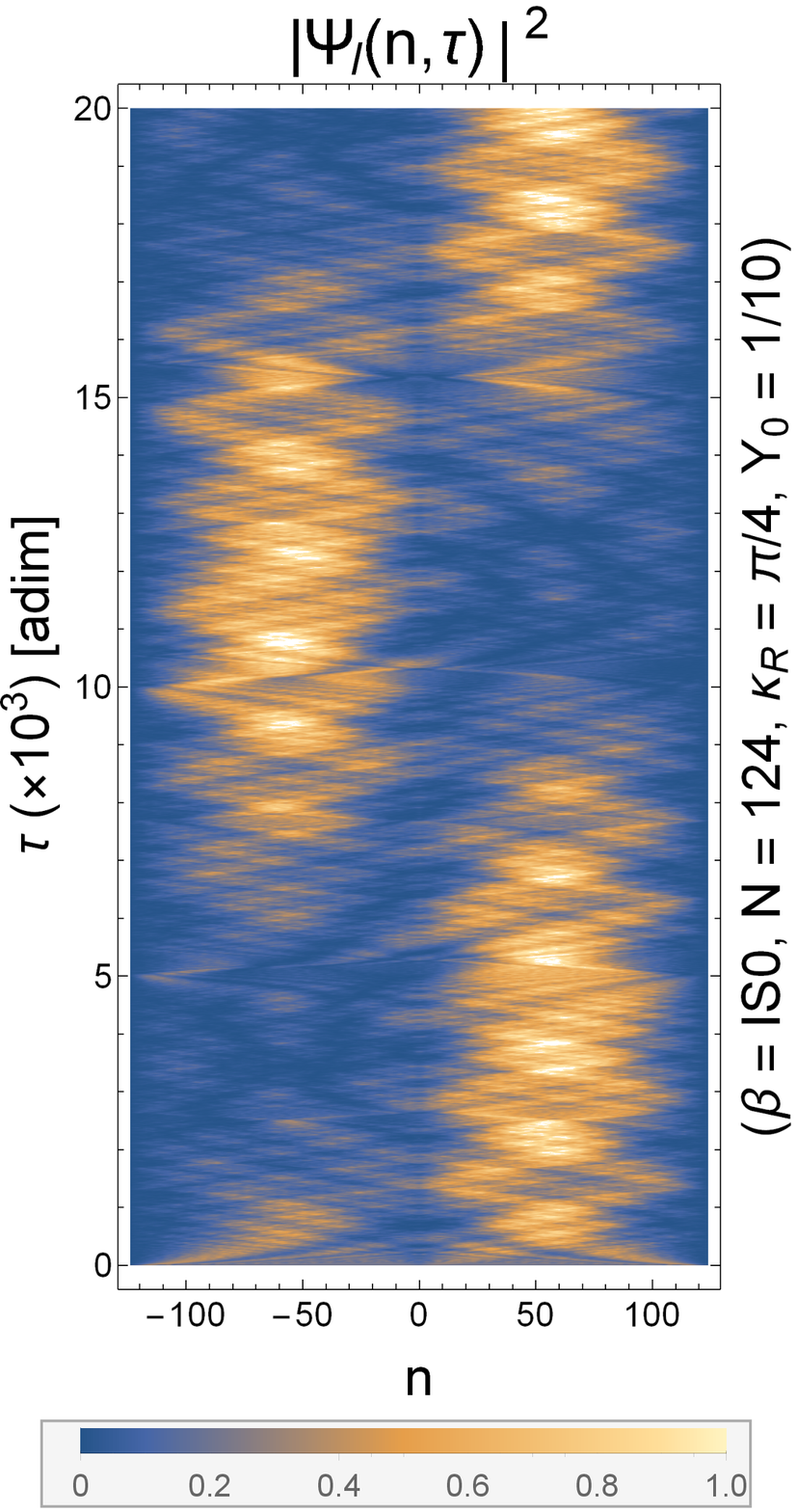}
				\caption{Evolution of a isoespectral wave packet in the interval
				$-124\leq n\leq 124$ (horizontal axis) at $\tau = 0$ (vertical axis),
				interacting with a Maxwellian potential located at $ n = 0$.}
			\label{fig:wBoltzPlotIS0-SG124-BI4-VG10.png}
		\end{figure}%
		
		An example of evolution is shown in fig.~\ref{fig:EntropyE-SG124-BI4-VG10.png}.
		The system size is $ 2N+1 = 249 $ sites, with scale parameter $a=L/2N$,
		the rescaled potential intensity is $\Upsilon_0 =\frac{1}{10}$ (=$ma^2V_0/\hbar^2$ [adim]),
		the reference momentum is $ \kappa_R = \pi/4 $ and $\beta =1/100$.		
		For relatively short times $\tau (\times10^{-3})\simeq 2$, a decrease of the entropy in \eqref{eq:ShEntropia} is appreciated.
		Then, between $\tau (\times10^{-3})\simeq 3$ to 9 the entropy increases which is explained by the natural wave expansion in each compartment,
		to decrease again at $\tau (\times10^{-3})\simeq 12$.

			Now we turn our attention to fig.~\ref{fig:Entropien-SG124-BI4-VG10.png}
		where we show a comparative plot of entropies by varying the temperature value.
		It is found that values $ 1/2 > \beta > 1/200 $ produce significant fluctuations for a potential intensity $\Upsilon_0 =\frac{1}{10}$.
		It should be stressed that for higher values of $\Upsilon_0$, the overall behavior shifts to larger values of beta.
		For very high temperatures, a highly disordered system in the energy basis has a tendency to fluctuate around its original entropic value
		(quasi-stationary behaviors), this implies that the effect is not strong in these cases.
		We have found, through these numerical results, that the role played by $\kappa_R$
		is partially decisive in the creation of box asymmetries in the evolution,
		as the intensity $\Upsilon_0$ is also important for small values of beta.
		However, we must stress that $\Upsilon_0$ cannot be taken as infinite, since all waves become trapped in such a case.
	
		In fig.~\ref{fig:ProbE-SG124-BI4-VG10.png} we can see asymmetries induced as time elapses,
		with must drastic effects occurring around $\tau (\times10^{-3})\simeq 8.5$
		where the difference between left and right probabilities (occupation) is large.
		Note that the effect is recurrent for larger times.
		In addition, the entropy has a minimum when the probability has a maximal rate of change with respect to time,
		implying that the Maxwellian potential operates reaching a quasi-stationary regime,
		where there is no exchange of densities but there is entropic rise.

			In fig.~\ref{fig:PsiHPsiE-SG124-BI4-VG10.png} we display lateral averages of the total energy as functions of time.
		We find asymmetries in both quantities: initially, the thermal wave is biased to the right.
		Then, between $\tau (\times10^{-3}) =$ 0 and 5 there is an expansion regime where there is thermalization.
		For $\tau (\times10^{-3}) >5$ the Maxwellian-potential's action enters the game and the waves are segregated again.
		These curves are compared with those of fig.~\ref{fig:PsiVPsiE-SG124-BI4-VG10.png},
		where indeed the average potential energy becomes negative for $\tau (\times10^{-3}) >5$,
		indicating that the particles exert work on the Maxwellian potential (see a global minimum of $\prom V$ at $\tau (\times10^{-3})\simeq 9$.
		In this setting, we conclude that our device operates well until the wave expansion allows an important interaction with $V$ at $\tau (\times10^{-3}) =5$ and after.
		For very large times, a regime with noisy collapse-and-revival behavior can be seen.
		
		In fig.~\ref{fig:wBoltzPlotE-SG124-BI4-VG10.png} we show a density plot for \eqref{eq:PsiBoltz}.					
		The Talbot effect induces a recurrence time in the quasi-temporal coordinate that will force the system to repeat its behavior.
		In this case, $\tau_\text{Talbot}\simeq 10(\times10^3)$, and
		for $\tau <\tau_\text{Talbot}$ there is an asymmetry that shows the efficient work of the Maxwellian potential.
		Subsequently the behavior is reversed between the compartments of the box.
		
		Another situation of interest is the uniform distribution, i.e. $\Psi_0^\text{I} = 1$,
		(this is denoted by $\beta=$IS0).
		For this case the entropy evolution is shown in fig.~\ref{fig:EntropyIS0-SG124-BI4-VG10.png}.
		Note that the entropy value oscillates, again reaching a minimum as the system evolves.
		A density plot of the wave function $\Psi_\text{I}(x,\tau)$ shown in fig.~\ref{fig:wBoltzPlotIS0-SG124-BI4-VG10.png},
		reveals that the wave is distributed asymmetrically due to terms that break parity explicitly as expected.

\section{Conclusions}
	We have dealt with an irreversible problem in time and space. In particular,
	we have reported a new asymmetrical Green's function in closed form pertaining to irreversible systems,
	not found in standard references \cite{GroscheSteiner1998}.
	The meromorphic structure of such a solution has been docile enough to allow
	proper identification of energy ranges where a Maxwellian sorting device is effective.
	In this way, we have identified how through a Fourier semi-transform
	the propagator of a real problem will be perturbed due to irreversibility.
	The symmetry breaking is located in a special term in the Green's function,
	whose pole is related with the reference energy at which a demon operates.

	Afterwards, a dynamical model for a system that splits an ensemble of waves
	representing independent particles has been proposed and successfully studied.
	Our description has been possible via a Hamiltonian operator given by \eqref{eq:HamV}
	and the irreversible potential in \eqref{eq:VactSgn}.
	The system works with a reference momentum that decides how two subsystems,
	with different temperatures, are distributed in each compartment of the cavity.
	The outcome is reminiscent of the classical demon's action shown in fig.~\ref{fig:EFEnsamble},
	as we have confirmed by analyzing wave dynamics in fig.~\ref{fig:wBoltzPlotE-SG124-BI4-VG10.png}.
		
	As an interesting result, the undulatory version of Maxwell's demon contains --in its evolution--
	the interference structure of Talbot (quantum) carpets in time domain.
	
	The reader familiarized with Jacobi theta functions may find in our interference patterns the typical trajectories of constant theta value that appear in many other applications,
	including factorization of natural numbers using Gaussian sums \cite{Schleich2007}.	
	For long times, a structure of collapses and revivals can be distinguished.
	This structure displays the expected spatial asymmetries for limited periods of time associated with Talbot lengths.
	There is no true thermalization (as opposed to the classical process) because of such revivals.
	A number of quantities and their time behavior support our conclusions in connection with irreversibility and the apparent entropy decrease.
	Indeed, with Shannon's definition for a basis-dependent disorder function (in energy states)
	we observe regimes where ordered configurations are established as time elapses.
	Also, densities and average energies at each compartment were studied.
	(Fig.~\ref{fig:PsiHPsiE-SG124-BI4-VG10.png} is unmistakable in this respect.)
	As mentioned in the introduction,
	our approach to irreversibility can be applied to many types of waves.
	Particular attention should be paid to electromagnetic cavities,
	since non-hermitian wave operators with odd space parity emerge naturally in dielectric media.
	Numerical implementations are left for future work.


\appendix

\section{\label{ap:a} Green's Function for a Dirac-delta Potential}
	For clarity and completeness to the case at hand in \eqref{eq:GreenVxp},
	we include the procedure to obtain the Green's function for the case of a $\delta(x)$-potential;
	such result can also be found at \citet{Blinder1988}, \citet{GroscheSteiner1998} and \citet{MoshinskySadurni2007}.
	To our knowledge, there is no prior reference to this result, although the 1D $\delta(x)$-potential appears in many textbooks.
	We start with
	\begin{equation}
		\Bparen{\hat{H}-E +V_0\delta(\hatx)} \hat G_\delta = \mathbb{I}.
	\end{equation}
	Multiplying from left with $\hat G_0$ and computing it in the position basis $x$
	\begin{equation}
		G_\delta(x,x',E) + V_0\bra{x} \hat G_0\delta(\hatx) \hat G_\delta\ket{x'} = G_0(x,x',E),
	\end{equation}
	where it has been used that
	\begin{equation}
		\Bparen{\hat{H}-E} \hat G_0 = \mathbb{I},
		\quad\text{and}\quad \bra{x} \hat G_0 \ket{x'} = G_0(x,x',E).
	\end{equation}	
	Inserting a continuous complete set, obtains
	\begin{align}\label{eq:GreenDeltaInt}\begin{split}
		G_\delta(x,x',E) &+ V_0\int dx'' G_0(x,x'',E) \delta(x'') G_\delta(x'',x',E) \\
			&= G_0(x,x',E).
	\end{split}\end{align}
	The above expression is evaluated at $x=0$, yielding a functional equation
	\begin{equation}
		G_\delta(0,x',E) = \frac{G_0(0,x',E)}{1+V_0 G_0(0,0,E)},
	\end{equation}
	which finally, when introduced in \eqref{eq:GreenDeltaInt}, obtains
	\begin{equation}\label{eq:GreenDelta}
		G_\delta(x,x',E) = G_0(x,x',E) - \frac{V_0G_0(x,0,E)G_0(0,x',E)}{1+V_0 G_0(0,0,E)}.
	\end{equation}

\section{\label{ap:b} Sine--integral Approximation and Meromorphic Structure}	
	In order to analyze the obtained Green's function,
	some integrals can be approximated. In particular,
	for a container, the terms to be obtained are
	\paragraph{The Fourier transform $\widetilde V(y)$}
	\begin{align}\begin{split}
		\widetilde V( y) &= \frac{1}{2\pi}\int dp\,e^{- ip(y+i\epsilon)}V(p) \\
			&= \frac{1}{2\pi}\left(\int_{-\infty}^{-P_R} + \int_{0}^{P_R} \right)dp\, e^{- ip(y+i\epsilon)} \\
			&\overset{\epsilon\rightarrow 0}{=} \frac{ 1}{2i\pi y}\Bparen{1-2\cos P_R y}.
	\end{split}\end{align}
	
	\paragraph{Integral $P_1^\text{C}(x',E)$}
	\begin{align}\begin{split}
		&P_1^\text{C}(x',E) = \int_{-L/2}^{L/2} dx\,\widetilde V(x)G_0^\text{C}(x,x',E) \\
			&= -\frac{2}{i\pi L}
			\sum_{n=1}\frac{\Si\bparen{\xi_+}-\Si\bparen{\xi_-}-\Si(n\pi)}{E_{2n}-E}\sin(\kappa_{2n}x'),
	\end{split}\end{align}
	where $\xi_\pm = \bparen{P_R\pm\kappa_{2n}}\frac{L}{2}$.
	
	\paragraph{Integral $Q_1^\text{C}(E)$}
	\begin{align}\begin{split}
		Q_1^\text{C}(E) &= \int_{-L/2}^{L/2} dx\,\widetilde V(x)\int_{-L/2}^{L/2} dy\,G_0^\text{C}(x,y,E)\widetilde V(-y)\\
			&=\frac{2}{\pi^2L}\sum_{n=1} \frac{\Bparen{\Si\bparen{\xi_+}-\Si\bparen{\xi_-}-\Si(n\pi) }^2}{E_{2n}-E}.
	\end{split}\end{align}
	
	\paragraph{Integral $P_2^\text{C}(x,E)$}
	\begin{align}\begin{split}
		P_2^\text{C}(x,E) &= \int dy\,G_0^\text{C}(x,y,E)\widetilde V(-y) \\
			&= \frac{2}{i\pi L}
			\sum_{n=1}\frac{\Si\bparen{\xi_+}-\Si\bparen{\xi_-}-\Si(n\pi)}{E_{2n}-E}\sin(\kappa_{2n}x) \\
			&= - P_1^\text{C}(x,E).
	\end{split}\end{align}
	
	Given the behavior of the function $\Si$, the following approximation can be made:
	The argument is written in the form
	\begin{align}\begin{autobreak}
		\Si\bparen{n\pi+a} - \Si\bparen{n\pi-a}
		= \Si\bparen{\pi(n+\pen{a/\pi})+\pi\epsilon}
		- \Si\bparen{\pi(n-\pen{a/\pi})-\pi\epsilon}
	\end{autobreak}\end{align}
	where $\pen{a/\pi}$represents the integer part and $\epsilon$ the fractional part of $a/\pi$.
	Expanding around $\epsilon=0$,
	\begin{align}\begin{split}
		&\Si\bparen{n\pi+a} - \Si\bparen{n\pi-a} \simeq \\
		&\Si\bparen{\pi(n+\pen{a/\pi})} - \Si\bparen{\pi(n-\pen{a/\pi})} \\
		&\qquad +
		\epsilon\left(\frac{\sin\bparen{\pi(n+\pen{a/\pi})}}{\bparen{n+\pen{a/\pi}}} + \frac{\sin\bparen{\pi(n-\pen{a/\pi})}}{\bparen{n-\pen{a/\pi}}}\right) \\
		&\simeq \frac{\pi}{2}-\frac{\pi}{2}\Theta\bparen{n-\pen{a/\pi}}+\frac{\pi}{2}\Theta\bparen{\pen{a/\pi}-n}+\pi\epsilon\,\delta_{n,\pen{a/\pi}}.
	\end{split}\end{align}
	(Note: Given the original function, the step function is zero for the case $n=\pen{a/\pi}$.)		
	Thus, it follows that
	\begin{align}\begin{split}
		\sum_{n=1}^\infty &\frac{\Si\bparen{\xi_+}-\Si\bparen{\xi_-}}{E_{2n}-E}\\
			&\simeq \sum_{n=1}^{\pen{a/\pi}-1}\frac{\pi}{E_{2n}-E} + \frac{\pi}{2}\sum_{n=1}^\infty\frac{\delta_{n,\pen{a/\pi}}}{E_{2n}-E},
	\end{split}\end{align}
	and
	\begin{align}\begin{split}
		\sum_{n=1}^\infty\frac{\Si\bparen{n\pi}}{E_{2n}-E} \simeq \frac{\pi}{2}\sum_{n=1}^\infty\frac{1}{E_{2n}-E},
	\end{split}\end{align}
	while
	\begin{align}\begin{split}
		\sum_{n=1}^\infty &\frac{\Bparen{\Si\bparen{\xi_+}-\Si\bparen{\xi_-}-\Si(n\pi) }^2}{E_{2n}-E}\\
			&\simeq \frac{\pi^2}{4}\sum_{n=1}^\infty\frac{1}{E_{2n}-E}
			- \frac{\pi^2}{4}\sum_{n=1}^\infty \frac{\delta_{n,\pen{a/\pi}}}{E_{2n}-E}.
	\end{split}\end{align}
	


%

\end{document}